\documentclass[conference]{IEEEtran}
\IEEEoverridecommandlockouts
\usepackage{cite}
\usepackage{nameref,hyperref}
\usepackage{amsmath,amssymb,amsfonts}
\usepackage{algorithmic}
\usepackage{graphicx}
\usepackage{textcomp}
\usepackage{xcolor}
\def\BibTeX{{\rm B\kern-.05em{\sc i\kern-.025em b}\kern-.08em
    T\kern-.1667em\lower.7ex\hbox{E}\kern-.125emX}}
\begin{document}

\title{VREd: A Virtual Reality-Based Classroom for Online Education Using Unity3D WebGL}

\author{\IEEEauthorblockN{ Ratun Rahman* }
\IEEEauthorblockA{\textit{Software Engineering} \\
\textit{Islamic University of Technology}\\
Gazipur, Bangladesh \\
ratunrahman@iut-dhaka.edu}
\and
\IEEEauthorblockN{ Md Rafid Islam}
\IEEEauthorblockA{\textit{Software Engineering} \\
\textit{Islamic University of Technology}\\
Gazipur, Bangladesh \\@article {Schlegel2016,
rafidislam@iut-dhaka.edu}
}
}
\maketitle

\begin{abstract}
Virtual reality is the way of the future. The use of virtual reality is expanding over time across all sectors, from the entertainment industry to the military and space. VREd is a similar concept where a virtual reality-based classroom is used for online education where the user will have better interaction and more control. Unity3D and WebGL software have been used for implementation. Students or learners accustomed to contemporary technologies may find the traditional educational system unappealing because of its flaws. Incorporating the latest technologies can increase the curiosity and learning abilities of students. The system architecture of VREd is similar to that of an actual classroom, allowing both students and teachers to access all of the course materials and interact with one another using only an internet connection. The environment and the background are also customizable. Therefore, all the users can comfortably use the system and feel at home. We can create an effective educational system that raises educational quality by utilizing virtual reality.
\end{abstract}


\section{Introduction}
Virtual reality has revolutionized the world that we see today. It has changed our perspectives on reality and introduced us to a new world that can be created, modified, and decorated by our imagination \cite{yoh2001reality}. Big tech companies like Apple, Meta, Samsung, etc. are investing billions of dollars to create a virtual world \cite{dwivedi2022metaverse}. Many researchers are also trying to find a new way to implement this technology in real life to transform the way of our life including education, training, medical instrument, military, simulation, industry, etc. \cite{pantelidis2010reasons}. 
The Internet has made it possible to learn a lot of things online while sitting at home. People from all walks of life are drawn to online learning because of the flexible schedules and simpler access to educational resources. People can learn at their own pace for a lower cost and acquire or improve various technical skills \cite{alam2022platform}. Online educational or learning systems are in demand, and their proper implementation with up-to-date technologies is also very important \cite{xue2022online}.

Grand View Research estimates that the global virtual reality market was worth USD 21.83 billion in 2021 and that it is anticipated to grow at a CAGR of 15.0\% from 2022 to 2030. Over 50\% of the market's revenue in 2021 came from the commercial segment, which is expected to continue to dominate the market \cite{grandviewresearchImmersiveMedia}. Also, VR is increasingly being used in aerospace and defense, healthcare, and various enterprises \cite{makransky2022benefits}. Due to the increasing demand for VR technology in the gaming and entertainment industries, the commercial sector is anticipated to experience significant growth \cite{safikhani2022immersive}.

This project includes a game-based approach that will make VREd both easier to understand and more fun to use. Multiple people can also access it at the same time and interact with each other. Therefore, it will eventually increase social interactions and human connections, which are major issues for online education \cite{gallace2022social}. 

The aim of the project is to introduce a unique and interactive way to learn and conduct educational services. This includes a virtual-reality-based application created in Unity3D and WebGL that can be hosted in the cloud across multiple platforms for easy internet access. This will be useful for not only students and teachers in institutions but also in online meetings and conferences, for people with disabilities, and overall for people who are interested in seeing modern technologies used properly in education. 

The rest of the paper is organized as follows: Section 2 of the paper contains a literature review discussing the core concepts and some other related works. In section 3, the research methodology was discussed including method and environment setup. In Section 4, the results and discussion were added, predominantly focusing on the benefits and drawbacks of a virtual reality-based classroom. Finally, in Section 5, the conclusion was added.

\section{Literature Review}
This section of the paper will be focused on introducing the concept of several core components and a brief history of similar research.

\subsection{Background Study}
\subsubsection{Virtual reality}
Interacting with synthetic, three-dimensional visual, or other sensory environments using computer modeling and simulation is referred to as "virtual reality". Users of VR applications are immersed in a computer-generated world that appears and feels real. This is accomplished by utilizing devices that can be worn as bodysuits, headsets, goggles, or gloves and that transmit and receive information \cite{rauschnabel2022xr}.

Our senses are the primary source of information about our surroundings, and our whole perception of reality is just a synthesis of the sensory data we receive and the cognitive processes we use to make sense of it \cite{nguyen2022human}. So it seems obvious that if our senses can be duped into receiving false information, our experience of reality will also be influenced. We would be shown a representation of reality that isn't actually there, but from our vantage point, it would appear to be realistic and something that can be described as virtual reality. Virtual reality can be summed up as the exposure of our senses to a digitally created environment that we may interact with in some way \cite{melo2022immersive}.

\subsubsection{Online Education}
Online education is a versatile method of delivering instruction that includes all forms of learnings that happens online \cite{lamsal2022exploring}. Students will not be required to physically attend their institution or school, and the full educational process would take place via the internet. The subjects and disciplines covered by online education are actually quite diverse. Nowadays, the internet is widely used and has enabled numerous developments. It has been used for educational purposes in quite effective manner as well \cite{li2022shift}. Other than specific courses or subjects that require hands-on lab experience, it is possible to learn most of the topics or subjects online.

An online educational program can be conducted through various methods, like video lectures, online assignments, and assessments. Video lectures can be delivered to a live audience or they may be pre-recorded \cite{greenhow2022foundations}. As technology keeps evolving and the internet becomes more accessible, it can be said with certainty that the scope of online education will be broadened as well. It remains to be seen how much more efficient the online education system will be through the use of various emerging technologies \cite{salas2022student}. 

\subsubsection{Unity}
Unity is a professional-level video game engine. It is used to make games like Rust, Assassin's Creed, and Superhot VR, as well as countless other games across different platforms. So, as we can see, unity is used to create games on platforms like Windows, Linux, Android, iOS, PS4, Xbox, or even VR \cite{xie2012research}. This is why this research has used Unity to create this virtual educational platform. Unity and its built-in functionalities can be used to create a video game environment and gameplay by simply telling them what to do. It is easier to use than most game engines but can perform all the functions. So, Unity has a perfect balance between sophistication and ease of use \cite{wang2010new}.

\subsubsection{WebGL}
WebGL is a simple, high-performance interactive 2D and 3D graphics programming interface for javascript applications without the use of plugins in any web browser. Unity has always used webGL API and its cross-platform component to create games that can be played in a browser. It helps the user visualize and perform in real-time by handling complex graphics and interactions. WebVR and WebAR are new additions to the WebGL component \cite{angel2014interactive}.

\subsection{Related Works}
Jaron Lanier, who is the founder of VPL Research, first used the word "virtual reality" in the mid-1980s \cite{zimmerman1986hand}. In the late 20th century, virtual reality was mainly used in flight simulation, industrial design, and military training \cite{liu2018virtual}. Virtual reality got its first big attention in 2012, and ever since then, its market has increased significantly. By 2016, all the big companies like Google, Amazon, Facebook, Microsoft, Samsung, Sony, etc. had a specialist group for AR and VR research \cite{wiredUntoldStory}. The pandemic in 2020 has skyrocketed the usage of virtual reality, and Grand View Research has stated that the market for VR will rise to 62.1 billion USD in 2027 \cite{grandviewresearchImmersiveMedia}.

There had been research carried out on virtual classrooms as well, especially during the COVID restriction when everything was shut down, including the education platform. In 1989, the first fully online course was provided by the University of Phoenix \cite{kentnor2015distance}. The virtual classroom: learning without limits via computer networks by SR Hiltz has talked about the benefits of virtual classrooms \cite{hiltz1994virtual}. However, it is not VR-based but more like an online classroom. VS Pantelidis in his Virtual reality in the classroom talked about the concept of virtual reality in the classroom in 1993. Liou, W.K. and Chang, C.Y. talked about VR for science in virtual reality classrooms applied to science education \cite{liu2020effects}. A review of the uses of virtual reality in engineering education by di Lanzo, J.A., Valentine, A., Sohel, F., Yapp, A.Y., Muparadzi, K.C. and Abdelmalek, M. talked about the benefits of VR in the classroom \cite{di2020review}. Are we ready for virtual reality in K–12 classrooms? by Araiza-Alba, P., Keane, T. and Kaufman, J. discussed the use of virtual reality for teaching a  younger generation \cite{araiza2022we}. There has also been a lot of research on virtual reality-based classrooms in recent years, especially in hand gesture recognition control \cite{islam2022nfs}.

\section{Research Methodology}
\subsection{Method}
A virtual classroom should contain all the basic functionalities of a classroom. Fig. 1 demonstrated a system architecture for a virtual classroom.
\begin{figure}[htbp]
\centerline{\includegraphics[width=0.5\textwidth]{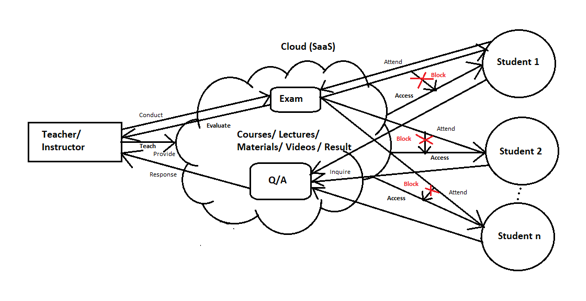}}
\caption{System Architecture for Virtual Classroom}
\label{fig1}
\end{figure}

As we can see, a teacher or instructor only requires an internet connection to provide service to the students. Similarly, a student also only needs an internet service to collect the materials. As a result, it simplifies the complexity of a virtual environment. Not only this, but the students can also access the materials at any available time to learn more effectively and efficiently. Therefore, cloud-based applications, more specifically SaaS-based applications can greatly impact the learning process of a student. Students can also ask questions and make inquiries at any time, and the teachers can also provide them at their respective times. During the exam, the students can also be blocked from accessing the materials or use some for open-book questions, which will provide better facilities and implementation. Therefore, by establishing all the core components correctly, a virtual classroom can be used as an alternative to an actual classroom.

\subsection{Environment}
It is important to set up a relevant environment for virtual classrooms as it can affect the study and concentration of students \cite{falloon2011making}. A very simple yet elegant classroom environment has been shown in fig. 2. As it is in a virtual environment, the seats, background scenes, and spaces can be rearranged and readjusted. Students can either pick a seat for themselves or they can be fixed by the authority as well.

\begin{figure}[htbp]
\centerline{\includegraphics[width=0.5\textwidth]{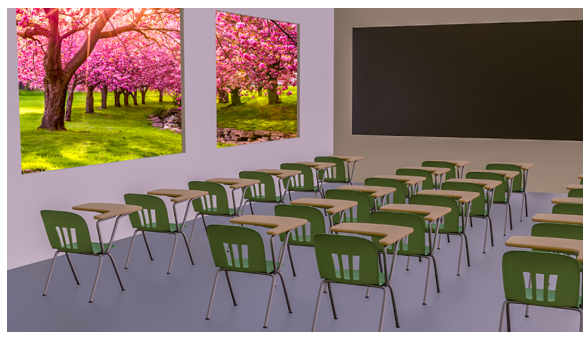}}
\caption{Classroom View}
\label{fig2}
\end{figure}

Fig. 3 demonstrates the desk or student view. After students are designed to sit in a particular place, their view should be similar to this. Students can also adjust their camera position, change the background, and increase visibility with the custom setup. This will enable the students to feel comfortable and at home.

\begin{figure}[htbp]
\centerline{\includegraphics[width=0.5\textwidth]{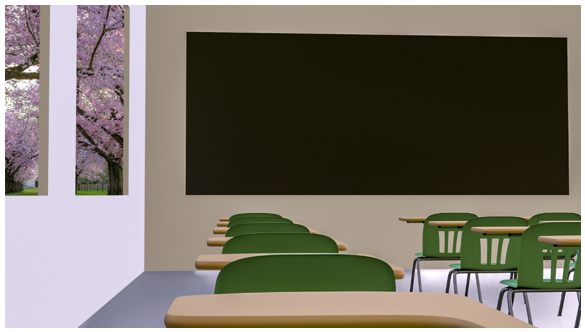}}
\caption{Desk/Student View}
\label{fig3}
\end{figure}

Fig. 4 displays the teacher's or instructor’s view. It is essential that the teacher can have a wider view of every student and get any queries from them. So, a teacher can alternate the view of the blackboard, mute or unmute any students, and also all other basic functionalities like changing the background, controlling the visibility, etc.

\begin{figure}[htbp]
\centerline{\includegraphics[width=0.5\textwidth]{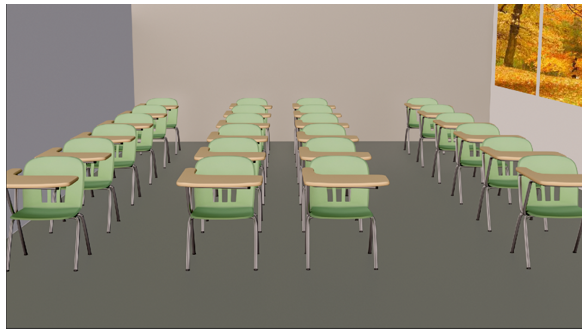}}
\caption{Teacher/Instructor View}
\label{fig4}
\end{figure}

\section{Result and Discussion}
This research contains a VR-based classroom with adjustable camera position, changeable backgrounds, and visibility control, which will enable the instructor and the learners to use the system with ease. Both benefits and disadvantages apply to this VR-based system.

Things that cannot be visualized in a traditional classroom or educational setting will be possible in a virtual reality setting. It is more interesting to watch and experience something than to read about it. There are various fields of study where the utilization of VR will result in a significant improvement in quality. Language barriers in education can also be overcome using VR technology. Various disciplines of engineering require field training, and VR can enable that with efficiency, which will save time. Since VR technologies will create an imaginary world to experience different realities, they will have fewer limitations and will aid in making experienced professionals with the lowest risk. The VR-based classroom will give the impression that students are in close proximity, interacting with each other, which may help reduce some of the negative effects caused by VR in terms of human relationships and social interactions. Overall, VR can be used to create an immersive educational curriculum that will have a better impact on students by making the entire learning procedure memorable and interactive.    

The technical expertise needed to use VR effectively would be its main drawback. Trained and efficient technical personnel are required to make use of VR technology. There can be software issues or issues with functionality that will require technical staff to fix and maintain, and even with efficient staff, it can be a major inconvenience. Setting up an infrastructure for a virtual classroom can be a real challenge as well. There can also be a lack of flexibility, depending on how the software has been programmed. Transitioning into a new environment is also very difficult \cite{islam2022survey}. Billion-dollar investments would be needed for the widespread use of VR technologies because they can be very expensive to implement, use, and maintain. Addiction to VR may be harmful to human psychology and deteriorate interpersonal relationships \cite{el2019virtual}.

\section{Conclusion}
The vast majority of the population is still unfamiliar with VR-based technologies, let alone a classroom that utilizes VR for educational purposes. But VR has the potential to make the learning process intriguing and advance the education system. Modern technological advancements and devices will make it simpler to adapt to VR and help tackle some of the existing issues. Still, there is a lot of work to be done if the aim is to properly make use of virtual reality in classrooms or education systems and take full advantage of it.

\section*{Acknowledgment}

This research received no external funding.

\bibliography{paper}

\bibliographystyle{ieeetr}

\end{document}